\documentclass[10pt,a4paper]{article}
\usepackage[latin1]{inputenc}
\usepackage{latexsym}
\usepackage{amsmath,amsfonts}
\usepackage{amsbsy}
\usepackage{mathrsfs}
\usepackage{color}
\usepackage{psfrag}
\usepackage{enumerate}
\usepackage{amsmath,amssymb,calc,amsfonts}
\usepackage{latexsym}
\usepackage[colorlinks=true,linkcolor=blue,urlcolor=blue,citecolor=red]{hyperref}
\usepackage{amsmath}
\usepackage{amsfonts}
\usepackage{amssymb}

\usepackage{latexsym}
\usepackage{amsfonts}
\usepackage{t1enc}
\usepackage{times}
\usepackage{xmpmulti}
\usepackage{color}
\usepackage{bm}
\usepackage{hyperref}

\font\tenscr=rsfs10 scaled1100
\font\sevenscr=rsfs7 
\font\fivescr=rsfs5 
\skewchar\tenscr='177
\skewchar\sevenscr='177
\skewchar\fivescr='177
\newfam\scrfam
\textfont\scrfam=\tenscr
\scriptfont\scrfam=\sevenscr
\scriptscriptfont\scrfam=\fivescr

\def\scri{{\fam\scrfam I}}

\begin{document}

\title{Spin and Center of mass comparison between the PN approach and the asymptotic formulation}
\date{\today}

\author{
\small C. N. Kozameh \\
\em \small FaMAF, Universidad Nacional de C\'ordoba\\
\normalsize \em \small 5000, C\'ordoba, Argentina \\
\normalsize \em \small kozameh@famaf.unc.edu.ar\\
\\
\small J. I. Nieva\\
\normalsize \em \small FaMAF, Universidad Nacional de C\'ordoba\\
\normalsize \em \small 5000, C\'ordoba, Argentina \\
\normalsize \em \small jnieva@famaf.unc.edu.ar
\\
\\
\small G. D. Quiroga\\
\normalsize \em \small GIRG, Escuela de F\'isica, Universidad Industrial de Santander\\
\normalsize \em \small A.A. 678, Bucaramanga, Colombia \\
\normalsize \em \small gonzalo.quiroga1@correo.uis.edu.co
}
\maketitle

\begin{abstract}
In this work we analyse the similarities and differences between the equations of motion for the center of mass and intrinsic angular momentum for isolated sources of gravitational radiation obtained by two different formulations. One approach is based on the asymptotic formulation of the GR whereas the other relies on Post-Newtonian methods. Several conclusions are obtained which could be useful for further developments in both approaches.
\end{abstract}

\maketitle

\section{Introduction}

The recent detections of gravitational waves made by LIGO \cite{LIGO2016,LIGO2016_2,LIGO2017,LIGO2017_2} have increased the interest in the study of binary systems and in the detection and characterization of the gravitational radiation emitted by these compact sources. In these observatories, the initial stage of the data analysis begins with the filtering of the measured signal. To improve the signal-to-noise ratio of the detector, the data output is compared with a bank of templates that represent the best theoretical predictions for the expected signals. The theoretical models that are used to construct these templates are based on Post-Newtonian (PN) methods which link the dynamical variables of the system to the emitted gravitational radiation in the non relativistic stage of the coalescence.

For these compact sources, it is very important to define the notion of center of mass and spin since the energy and momentum carried away by the gravitational wave induce a recoil to the center of mass of the coalesced binary. Likewise, the spin of the resulting black hole or neutron star depends on the emitted gravitational wave. Although, care must be taken to define these notions, in the PN approximation one starts with a Newtonian definition, since it is assumed that when the compact objects are far away the gravitational radiation is negligible and the system is well described by Newtonian orbiting particles. As the sources get closer one redefines these variables using the available Hamiltonian for the required approximation. However, in the very energetic regime a full GR definition should be given. Otherwise, one is at risk of obtaining erroneous results for the final recoil speed or final spin of the resulting black hole or neutron star. The problem lies in the impossibility of defining locally these variables since the gravitational radiation gives a vanishing contribution to the stress energy tensor, though it carries away energy, momentum and angular momentum.

On the other hand, using the notion of asymptotic flatness together with the inclusion of a 3-dim null boundary, called Null Infinity, one defines global variables for the isolated system like the Bondi mass $M_B$, linear momentum $P_B^i$ \cite{BBM}, and the mass dipole-angular momentum two-form $M_{\mu \nu}$. These global variables are constructed from suitable integrals at null infinity of the available radiative fields. This "Gaussian" approach yields physically meaningful flux laws for the above mentioned variables. This fact has been acknowledged in the PN approach and the flux laws derived for asymptotically flat spacetimes are used in the PN formalism\cite{BlanchetI}. Moreover, the  relationship between the local description of the motion of the sources and the Bondi mass, linear and angular momentum is computed at every stage of the approximation procedure\cite{BlanchetI}. Nevertheless, it is not an easy task in the PN approach to define the center of mass worldline, and relate its motion to the available global quantities defined at null infinity. Many authors define the center of mass velocity as  $V^i\equiv P_B^i/M_B$. However, in doing so one could be neglecting the contribution of the gravitational radiation to the Bondi momentum. (The analogous definition of total linear momentum for interacting charged particles explicitly contains the kinematical particle as well as the Maxwell field contribution, see eq. (33.6) in ref. \cite{landeau}). This in turn could give an erroneous result when computing the recoil velocity in a given coalescence problem. One should also mention that without an adequate definition of center of mass it is impossible to define the intrinsic angular momentum of the system.

In a recent work, a definition of center of mass and intrinsic angular momentum for isolated sources of gravitational radiation based on global quantities defined at null infinity was given and their time evolutions were derived \cite{kozameh2016center}. A key issue in the formulation is the use of a special set of Newman-Unti congruences that foliate null infinity as a one parameter families of cuts. Each foliation is associated with a worldline in a fiducial Minkowski space called observation space. It was shown that for one such foliation the associated mass dipole moment vanishes. Thus, the special worldline with vanishing mass dipole moment is called center of mass. Moreover, the angular momentum of this foliation is called intrinsic angular momentum. This formulation yields by construction a regular worldline and the evolution equations for the center of mass and spin are derived from the available Bondi evolution equations for the radiative fields at null infinity. The whole construction is global and
regular since by assumption all the radiative fields are regular at null infinity. A non trivial task in this formulation is to relate these global variables with the motion of sources in the spacetime and it is part of ongoing research. In this regard, a comparison between similar variables that are used in the PN and our approach should be of great help to obtain a robust approximation scheme in both formulations.

It is then the purpose of this work to compare the evolution equations for the center of mass and intrinsic angular momentum in both formalisms. The first result is promising: both formulations yield identical results if one only keeps the quadrupole mode of the radiative field (as we will see in the derived equations). This is somehow surprising since the PN approach is based on the motion of the sources and the asymptotic formulation is based on the behavior of the radiative fields at null infinity.  Using this result as a guideline we then compute the nontrivial deviation in both formulations. To do so, we extend our earlier work since the original derivation only kept quadrupole terms. We find that adding an octupolar contribution yields the first non trivial difference between the formalisms.
The slow motion approximation is also assumed in our approach since the center of mass do not acquire relativistic velocities as a result of the gravitational radiation emission. It is also necessary to compare our derivations with the PN results. As a result of this approximation spin-velocity terms will be neglected.

The paper is organized as follows. In Section 2 we give a summary of our previous results and some mathematical tools needed for our constructions. In particular, we introduce the dipole mass moment and total angular momentum vector for an isolated source coming from the Linkage integral. In section 3 we derive the main results obtaining the relationships between these global variables together with their time evolution. In section 4 we compare our evolution equations with those coming from the Post Newtonian formalism. Finally, we conclude this work with some remarks and conclusions about the PN approach and our asymptotic formulation.

\section{A brief summary of Asymptopia}
In this section, we briefly review some results derived within the framework of asymptotically flat spacetimes that will be useful for this work.

The notion of an asymptotically flat spacetime \cite{newman1980asymptotically}, the Newman-Penrose formalism \cite{newman1962approach}, and the notion of mass dipole/angular momentum introduced by the Winicour-Tamburino linkage \cite{tamburino1966gravitational} play a central role in our construction. A thorough review about these formalisms can be found in the following references \cite{newman1980asymptotically,quiroga2017asymptotic,LMN}.

We first introduce two sets of coordinates labeled by $(u_B,r_B, \zeta,\bar{\zeta})$, and $(u,r, \zeta,\bar{\zeta})$ to denote the Bondi and Newman-Unti (NU) coordinates respectively. In both sets, $(u_B,u)$ represents the Bondi and the Newman-Unti time. These coordinates label foliations of  cuts of $\scri^+$, the null boundary of the null infinity, and are used to identify the null surfaces that intersect null infinity at the corresponding cuts. One then introduce  affine parameters $r_B$ and $r$ along the null geodesics of the null surfaces now labelled as $u_B=const.$ and $u=const.$. Finally, $\zeta = e^{i\phi} \cot(\theta/2)$, is the complex stereographic coordinate labeling the null geodesics of each null surface. Associated with these coordinates one has also available the null tetrads,
\begin{eqnarray}
(l^a,n^a,m^a,\bar{m}^a), \label{baseB}\\
(l^{\ast a},n^{\ast a},m^{\ast a},\bar{m}^{\ast a}), \label{baseNU}
\end{eqnarray}
here the ${}^\ast$ denote the associated vectors with the NU system. The NU foliations determined by the condition $u = const.$ are related to those of Bondi through the transformations,
\begin{align}
  u_B &= Z(u,\zeta,\bar{\zeta}), \\
  r_B &= Z^{\prime} r,
\end{align}
where $Z$ is a real function, and $Z^{\prime}$ denotes the $\partial_u Z$. Moreover, these equations allow to establish a relation between the sets of vectors. These vectors, or tetrad of vectors, form a base of the spacetime, and the transformation law between these bases is given by the following equations,
\begin{align}
l_{a}^{\ast } &=\frac{1}{Z^{\prime }}[l_{a}-\frac{L}{r_{B}}\bar{m}_{a}-%
\frac{\bar{L}}{r_{B}}m_{a}+\frac{L\bar{L}}{r_{B}^{2}}n_{a}], \label{la}\\
n_{a}^{\ast } &=Z^{\prime }n_{a}, \label{na}\\
m_{a}^{\ast } &=m_{a}-\frac{L}{r_{B}}n_{a} \label{ma}, \\
\bar{m}_{a}^{\ast } &=\bar{m}_{a}-\frac{\bar{L}}{r_{B}}n_{a}, \label{mba}
\end{align}
where
\begin{equation}
L(u_{B},\zeta ,\bar{\zeta })=\eth Z(u,\zeta ,\bar{\zeta }).
\end{equation}
The way in which this function is chosen is one of the main inputs of this work. We demand that $Z$ satisfy the regularized null cone (RNC) cut equation \cite{kozameh2016center},
\begin{equation}\label{RNC}
\bar{\eth}^2 \eth^2 Z= \bar{\eth}^2 \sigma^{0}(Z,\zeta,\bar\zeta)+ {\eth}^2 \bar{\sigma}^{0}(Z,\zeta,\bar\zeta).
\end{equation}

A straightforward way to get this equation is to solve the linearized geodesic deviation equation for the future light cone from a point. It represents the Huygens part of the intersection of the future light cone from a given point of the spacetime with null infinity.
In previous works \cite{kozameh2016center,kozameh2012spin} we have discussed in detail about the RNC cut equation and we have shown how to obtain a NU foliation from the null cone cuts of null infinity. Extra details about the RNC cuts are given in \cite{bordcoch2016asymptotic}.One should also mention that the RNC cut equation coincides with the linearized L. Mason equation \cite{mason1995vacuum} obtained following a completely different approach,

Another useful variables are twelve complex quantities called ``Spin Coefficients'' and five complex scalars named ``Weyl Scalars''. These complex scalars are built from the Ricci rotation
coefficients and from the contraction of the null vectors with the Weyl tensor respectively. However, the most important scalars in our approach are introduced below,
\begin{align}
\psi_1 &\simeq \frac{\psi_1^0}{r_B^4},  \qquad \psi_1^{\ast} \simeq \frac{\psi_1^{\ast 0}}{r^4}, \\
\sigma &\simeq \frac{\sigma^0}{r_B^2},  \qquad \sigma^{\ast} \simeq \frac{\sigma^{\ast 0}}{r^2}.
\end{align}
Here the Weyl scalar ${\psi }_{1}^{0\ast}$ is constructed from the NU tetrad (\ref{baseNU}) while ${\psi }_{1}^{0}$  from the tetrad (\ref{baseB}).  The variables $\sigma^{0\ast}$ and $\sigma^{0}$ are respectively called the asymptotic NU and Bondi shears. These quantities are related by the following equations \cite{kozameh2016center,aronson1972coordinate},
\begin{align}
\frac{{\psi }_{1}^{0\ast}}{Z^{\prime 3}}&=[\psi _{1}^{0}-3L\psi _{2}^{0}+3L^{2}\psi_{3}^{0}-L^{3}\psi _{4}^{0}],\label{transformacion} \\
\frac{\sigma^{0\ast }}{Z^{\prime}}&=\sigma ^{0}-\eth^{2}Z.\label{sigma*}
\end{align}

For any stationary spacetimes, at a linearized level, the real and imaginary parts of $\psi _{1}^{0}$ capture the notion of the two-form that defines the dipole mass and angular momentum. Thus, for any asymptotically flat spacetimes a natural generalization of dipole mass moment-angular momentum tensor arise from the Winicour-Tamburino linkage \cite{tamburino1966gravitational} for a given a $u=const.$ null foliation, which can be either NU or Bondi. To obtain these components, it is quite convenient to define a complex vector  $D^{\ast}_i+\frac{\mathtt{i}}{c}J^{\ast}_i$ (see ref. \cite{kozameh2016center,lind1972equations} for extra details) as,
\begin{eqnarray}\label{DJNU}
D^{\ast i}+\frac{\mathtt{i}}{c}J^{\ast i} =-\frac{c^{2}}{12\sqrt{2}G}  \left[ \frac{2\psi _{1}^{0\ast}-2\sigma^{0\ast}\eth^{\ast} \bar{\sigma}^{0\ast}-\eth^{\ast}(\sigma ^{0\ast}\bar{\sigma}^{0\ast})}{Z^{\prime 3}}\right]^i.
\end{eqnarray}
Now, in a Bondi system the last equation take the form,
\begin{equation}\label{DJB}
D^{i}+ic^{-1}J^{i}=-\frac{c^{2}}{12\sqrt{2}G}\left[ 2\psi _{1}^{0}-2\sigma^{0}\eth \bar{\sigma}^{0}-\eth(\sigma ^{0}\bar{\sigma}^{0})\right]^i.
\end{equation}
It is possible to relate eq. (\ref{DJNU}) and (\ref{DJB}) just using the transformation law introduced before, see eqs. (\ref{transformacion}) and (\ref{sigma*}), to obtain the following equation,
\begin{eqnarray}
D^{\ast i}(u) &=&D^{i}(u_B)+\frac{3c^{2}}{6\sqrt{2}G}Re[\eth Z(\Psi -\eth ^{2}\bar{\sigma}^{0})+F]^{i}\label{Dexp} \\
J^{i\ast }(u) &=&J^{i}(u_B) +\frac{3c^{3}}{6\sqrt{2}G}Im[\eth Z(\Psi -\eth ^{2}\bar{\sigma}^{0})+F]^{i}\label{Jexp}
\end{eqnarray}
where the complex function $F$ is given by,
\begin{eqnarray}\label{F}
F&=&-\frac{1}{2}(\sigma ^{0}\eth \bar{\eth }^{2}Z+\eth ^{2}Z\eth \bar{\sigma}^{0}-\eth ^{2}Z\eth \bar{\eth }^{2}Z)\nonumber\\
&&-\frac{1}{6}(\bar{\sigma}^{0}\eth ^{3}Z+\bar{\eth }^{2}Z\eth \sigma ^{0}-\bar{\eth }^{2}Z\eth ^{3}Z).
\end{eqnarray}
Finally, we introduce the notion of Bondi mass and linear momentum, these equations are usually written as \cite{newman1980asymptotically}
\begin{eqnarray}
\left[\psi _{2}^{0}+\eth^{2}\bar{\sigma }^{0}+\sigma ^{0}\dot{\bar{\sigma}}^0\right]|_{\ell=0}&=&-\frac{2\sqrt{2}G}{c^{2}}M, \label{massM}\\
\left[\psi _{2}^{0}+\eth^{2}\bar{\sigma }^{0}+\sigma ^{0}\dot{\bar{\sigma}}^0\right]_{\ell=1}^i&=&-\frac{6G}{c^{3}}P^{i}. \label{momentoP}
\end{eqnarray}
The superscript $i$ denotes the three-vector associated with a tensorial spin-s decomposition as we see in the next section.

\section{Equations of motion for the center of mass and angular momentum}

\subsection{Approximations and assumptions} \label{supuestos}
We have previously defined the notion of mass dipole moment and angular momentum associated with a NU or Bondi congruence. In particular, eqs. (\ref{Dexp})-(\ref{Jexp}) give a relation between these variables. Introducing a tensorial spin-s spherical harmonics decomposition; $Y^0_0, Y^0_{1i}, Y^0_{2ij}$, etc.\cite{newman2005tensorial} and keeping up to quadrupole and octupole terms, we can expand the relevant scalars as,
\begin{eqnarray} \label{expan_tensorial}
\sigma^0 &=&\sigma ^{ij}(u_{B})Y_{2ij}^{2}(\zeta,\bar \zeta )+\sigma ^{ijk}(u_{B})Y_{3ijk}^{2}(\zeta,\bar \zeta ), \\
\psi _{1}^{0} &=&\psi _{1}^{0i}(u_{B})Y_{1i}^{1}(\zeta,\bar \zeta )+\psi_{1}^{0ij}(u_{B})Y_{2ij}^{1}(\zeta,\bar \zeta )+\psi_1^{0ijk}(u_B)Y_{3ijk}^1, \\
\Psi &=&-\frac{2\sqrt{2}G}{c^{2}}M-\frac{6G}{c^{3}}P^{i}Y_{1i}^{0}(\zeta,\bar \zeta )+\Psi^{ij}(u_B)Y^0_{2ij}(\zeta,\bar\zeta)\\
&+&\Psi^{ijk}(u_B)Y^0_{3ijk}(\zeta,\bar\zeta).\nonumber
\end{eqnarray}
The complex tensor $\sigma^{ij}$ ($\sigma^{ijk}$ ) represents the radiative quadrupole (octupole) contribution of the gravitational wave. The real and imaginary parts of $\sigma^{ij}$ ($\sigma^{ijk}$ ) are respectively called, the "electric" and "magnetic" parts.

Since the mass dipole moment should vanish at the center of mass position, the condition $D^*=0$ gives the position of the center of mass in a Bondi coordinate system by evaluating the r.h.s. of eq. (\ref{Dexp}).
Similarly, the angular momentum at the center of mass position $J^{*i}=S^i$ is, by definition, the spin or intrinsic angular momentum of the system. Finally, eq. (\ref{Jexp}) gives a relation between the spin and the total angular momentum which will be obtained explicitly in the following subsection.

\subsection{The center of mass and spin} \label{CoM Spin}

The center of mass worldline $X^{i}(u)$ is obtained from (\ref{Dexp}) by demanding that the l.h.s. vanishes on the $u=const.$ cut when $u_B=Z_1(u,\zeta,\bar \zeta)$ is inserted in the r.h.s. of the equation. Furthermore, since by assumption $X^{i}(u)$, $\sigma_R^{ij}(u)$, and $\sigma_R^{ijk}(u)$ are small, also we introduce the first order solution of the RNC cut (\ref{RNC}) as follows,
\begin{equation}\label{Z1}
Z_1=u+\delta u=u+\delta u,
\end{equation}
with
\begin{equation}\label{du}
\delta u=-\frac{1}{2}X^{i}(u)Y_{1i}^{0}+\frac{1}{12}\sigma _{R}^{ij}(u)Y_{2ij}^{0}+\frac{1}{60}\sigma _{R}^{ijk}(u)Y_{3ijk}^{0}
\end{equation}
and making a Taylor expansion of eqs. (\ref{Dexp}) and (\ref{Jexp}) up to first order in $\delta u$ we get,
\begin{equation}\label{1}
0=D^i+\frac{c^2}{6\sqrt{2}G}Re[(\eth\Psi-\eth^3\bar\sigma^0)\delta u]^i+\frac{3c^2}{6\sqrt{2}G}Re[(\Psi-\eth^2\bar\sigma^0)\eth\delta u+F]^i
\end{equation}
and
\begin{equation}\label{1}
S^i=J^i+\frac{c^3}{6\sqrt{2}G}Im[(\eth\Psi-\eth^3\bar\sigma^0)\delta u]^i+\frac{3c^3}{6\sqrt{2}G}Im[(\Psi-\eth^2\bar\sigma^0)\eth\delta u+F]^i,
\end{equation}
where $F$ is given by (\ref{F}).\

Now, using the definition of $\delta u$, $\Psi$, $\bar\sigma^0$, and considering only linear terms in $\delta u$ and $\delta u^{\prime}$ we obtain,
\begin{equation}
MX^i=D^i + \frac{8}{5\sqrt{2}c}\sigma_R^{ij}P^j\label{cm}.
\end{equation}
Also from eq. (\ref{1}) we can get the relation between the spin and the total angular momentum as follows,
\begin{eqnarray}\label{am}
J^i&=&S^i +\epsilon^{ijk}X^j P^k+\frac{137c^3}{168\sqrt{2}G}(\sigma_R^{ijk}\sigma_I^{jk}-\sigma_I^{ijk}\sigma_R^{jk}).\label{jotai}
\end{eqnarray}

\subsection{Dynamical Evolution} \label{DynEvo}

The time evolution of $D^i$ and $J^i$ can be obtained taking one time derivative of eq. (\ref{DJB}) togethers with the equation for $\dot{\psi}^0_1$ \cite{kozameh2016center}. Furthermore, the dynamical of the Bondi mass and momentum $P$ can be computed from the Bianchi identity for $\dot{\psi}_2^0$. These equations are given by,
\begin{eqnarray}
\dot{D}^{i}&=&P^{i}+\frac{3}{7}\frac{c^2}{\sqrt{2}G}\big[(\dot\sigma_R^{ijk}\sigma_R^{jk}-\sigma_R^{ijk}\dot\sigma_R^{jk})+(\dot\sigma_I^{ijk}\sigma_I^{jk}-\sigma_I^{ijk}\dot\sigma_I^{jk}) \big] ,\label{realpart}\\
\dot{J}^{i}&=&\frac{c^{3}}{5G}(\sigma _{R}^{kl}\dot{\sigma}_{R}^{jl}+\sigma _{I}^{kl}\dot{\sigma}_{I}^{jl})\epsilon ^{ijk}+\frac{9c^3}{7G}(\sigma_R^{klm}\dot\sigma_R^{jlm}+\sigma_I^{klm}\dot\sigma_I^{jlm})\epsilon^{ijk},\label{impart}\\
\dot{M}&=&-\frac{c}{10G}(\dot{\sigma}_{R}^{ij}\dot{\sigma}_{R}^{ij}+\dot{\sigma}_{I}^{ij}\dot{\sigma}_{I}^{ij})-\frac{3c}{7G}(\dot\sigma_R^{ijk}\dot\sigma_R^{ijk}+\dot\sigma_I^{ijk}\dot\sigma_I^{ijk}  ),\label{mpunto}\\
\dot{P}^{i} &=&\frac{2c^{2}}{15G}\dot{\sigma}_{R}^{jl}\dot{\sigma}%
_{I}^{kl}\epsilon ^{ijk}-\frac{\sqrt{2}c^2}{7G}(\dot\sigma_R^{jk}\dot\sigma_R^{ijk}+\dot\sigma_I^{jk}\dot\sigma_I^{ijk})+\frac{3c^2}{7G}\dot\sigma_R^{jlm}\dot\sigma_I^{klm}\epsilon_{ijk}. \label{ppunto}
\end{eqnarray}
These above equations are used to derive the equation of motion for the center of mass.

Starting from (\ref{cm}), and taking one time derivative it is posible to obtain the relation between the center of mass velocity and the scalars at null infinity. Considering up to quadratic terms, this equation reads,
\begin{equation}\label{momento}
M\dot{X}^{i}=P^{i}+\frac{8}{5\sqrt{2}c}\dot{\sigma}_{R}^{ij}P^{j}+ \frac{3c^2}{7\sqrt{2}G}[(\dot\sigma_R^{ijk}\sigma_R^{jk}-\sigma_R^{ijk}\dot\sigma_R^{jk})+(\dot\sigma_I^{ijk}\sigma_I^{jk}-\dot\sigma_I^{jk}\sigma_I^{ijk}) ].
\end{equation}
Finally, taking one more time derivative of (\ref{momento}) and considering up quadratic terms one obtains the equation of motion for the center of mass,
\begin{eqnarray}
M\ddot X^i-\frac{8M}{5\sqrt{2}c}\ddot\sigma_R^{ij}\dot X^j&=&\frac{2c^{2}}{15G}\dot{\sigma}_{R}^{jl}\dot{\sigma}_{I}^{kl}\epsilon ^{ijk}-\frac{\sqrt{2}c^2}{7G}(\dot\sigma_R^{jk}\dot\sigma_R^{ijk}+\dot\sigma_I^{jk}\dot\sigma_I^{ijk})\\
&+&\frac{3c^2}{7G}\dot\sigma_R^{jlm}\dot\sigma_I^{klm}\epsilon_{ijk}+\frac{3c^2}{7\sqrt{2}G}(\ddot\sigma_R^{ijk}\sigma_R^{jk}-\sigma_R^{ijk}\ddot\sigma_R^{jk})\nonumber\\
&+&\frac{3c^2}{7\sqrt{2}G}(\ddot\sigma_I^{ijk}\sigma_I^{jk}-\sigma_I^{ijk}\ddot\sigma_I^{jk}).\label{Xddot}\nonumber
\end{eqnarray}
Following the same steps for the angular momentum, we can write,
\begin{eqnarray}\label{Sdot}
\dot{S}^{i}&=&\dot J^i+\frac{137c^3}{168\sqrt{2}G}(\sigma_I^{jk}\sigma_R^{jki})^{\cdotp}-\frac{137c^3}{168\sqrt{2}G}(\sigma_I^{jki}\sigma_R^{jk})^{\cdotp}\nonumber\\
&=&\frac{c^{3}}{5G}(\sigma _{R}^{kl}\dot{\sigma}_{R}^{jl}+\sigma _{I}^{kl}\dot{\sigma}_{I}^{jl})\epsilon ^{ijk}+\frac{9c^3}{7G}(\sigma_R^{klm}\dot\sigma_R^{jlm}+\sigma_I^{klm}\dot\sigma_I^{jlm})\epsilon^{ijk}\\
&+&\frac{137c^3}{168\sqrt{2}G}(\sigma_I^{jk}\sigma_R^{jki})^{\cdotp}-\frac{137c^3}{168\sqrt{2}G}(\sigma_I^{jki}\sigma_R^{jk})^{\cdotp}\nonumber
\end{eqnarray}

\section{A Comparison with the Post Newtonian formalism}
In this section we compare our evolution equations with those coming from the Post Newtonian formalism.
The asymptotic formulation has exact equations of motion for the total Bondi mass, linear and angular momentum of the isolated system. Furthermore, there is a well defined procedure to first obtain the center of mass vector and spin and then derive their equations of motion. Although we have used a slow motion approximation and kept up to octupole contributions in a spherical harmonic decomposition,  our procedure can in principle be implemented for any order of approximation and  for arbitrary spherical harmonic contributions.  Since the main goal of this work  is to compare our results with those coming from the Post Newtonian formalism it is worth mentioning that in the Post Newtonian approach one builds up the loss due to gravitational radiation valid up to the level of approximation considered since a priori one does not have available an exact formula for the center of mass or intrinsic angular momentum. Thus, it is not an easy task to match orders of approximation in these apparently dissimilar
approaches to the emission of gravitational waves.

Nevertheless it is very useful to try to see whether or not they yield equivalent equations of motion for a compact source emitting gravitational radiation. A matching of the formulae could give a robust check for the formulations and the discrepancies should be useful to detect possible sources of errors in the formalisms.

We compare below the evolution equations for the total mass, momentum and angular momentum of a compact source of gravitational radiation. In both formalisms, a dot derivative means derivation with respect with the retarded time.\

The PN formalism also uses the Bondi radiative energy, linear and angular momentum loss available for asymptotically flat space times \cite{Blanchet2014,BlanchetQ},

\begin{eqnarray}
\dot E_{PN}=&-&\frac{G}{5c^5}U^{(1)ij}U^{(1)ij}-\frac{16G}{45c^7}V^{(1)ij}V^{(1)ij}-\frac{G}{189c^7}U^{(1)ijk}U^{(1)ijk}\nonumber\\
&-&\frac{G}{84c^9}V^{(1)ijk}V^{(1)ijk} \\
\dot P^i_{PN}=&-&\frac{2G}{63c^7}U^{(1)ijk}U^{(1)jk}+\frac{16G}{45c^7}\epsilon^{ijk}U^{(1)kl}V^{(1)jl}-\frac{4G}{63c^9}V^{(1)ijk}V^{(1)jk}\nonumber\\
&+&\frac{1G}{126c^9}\epsilon^{ijk}U^{(1)klm}V^{(1)jlm}\\
\dot S^i_{PN}=&-&\epsilon^{ijk}G \Big( \frac{1}{c^5} \frac{2}{5}U^{kl}U^{(1)jl}+\frac{1}{c^5}\frac{32}{45}V^{kl}V^{(1)jl}\nonumber\\
 &+&\frac{1}{c^7}\frac{1}{63}U^{klm}U^{(1)jlm} +\frac{1}{c^7}\frac{1}{28}V^{klm}V^{(1)jlm} \Big),
\end{eqnarray}

where in the above equations the quadrupole as well octupole terms have been included.

Since both formalisms use the same equation for these global variables, making the following identification of quadrupole and octupole terms
\begin{align}
 \sigma_R^{ij}&\rightarrow -\frac{\sqrt{2}G}{c^3}U^{ij}\\
 \sigma_I^{ij}&\rightarrow \frac{4\sqrt{2}G}{3c^4}V^{ij}\\
 \sigma_R^{ijk}&\rightarrow -\frac{G}{9c^4}U^{ijk}\\
 \sigma_I^{ijk}&\rightarrow \frac{G}{6c^5}V^{ijk}
\end{align}
one obtains identical expressions for the mass and linear momentum loss formulae. This is not surprising since, as we said before, both approaches use the same Bondi flux equations. However, as we will see below, this does not imply that the acceleration or the time evolution of the center of mass are identical in both approaches.

It is worth noting that in the PN formalism most, if not all, of the results are obtained in the center of mass frame. In order to compare the acceleration of the center of mass in both approaches we have to find the appropriate Bondi frame such that at a given initial time $u_0$ the system was not radiating and,
\begin{equation}
X_0^i=0, \qquad \dot{X}_0^i=V_0^i=0
\end{equation}
and therefore
\begin{equation}
P_0^i=0,
\end{equation}
i.e., the initial Bondi momentum vanishes in our setup. Keeping up to quadratic terms in the radiative shear we get directlly from (\ref{momento}),
\begin{equation}
M V^i= P^i+\frac{3c^2}{7\sqrt{2}G} [(\dot\sigma_R^{ijk}\sigma_R^{jk}-\sigma_R^{ijk}\dot\sigma_R^{jk})+(\dot\sigma_I^{ijk}\sigma_I^{jk}-\sigma_I^{ijk}\dot\sigma_I^{jk})],
\end{equation}
from which we obtain
\begin{equation}
V^i= V_{PN}^i+  \frac{3c^2}{7M\sqrt{2}G} [(\dot\sigma_R^{ijk}\sigma_R^{jk}-\sigma_R^{ijk}\dot\sigma_R^{jk})+(\dot\sigma_I^{ijk}\sigma_I^{jk}-\sigma_I^{ijk}\dot\sigma_I^{jk})].
\end{equation}
In the above equation we have used the recoil velocity of the center of mass that is defined in the PN formalism as $P_B^i/M_B$. As one can see, the two velocities differ by octupole (and higher) terms.

Integrating again yields a relation between the center de mass positions in both formalism,
\begin{equation}
X^i = X^i_{PN}+\frac{3c^2}{7M\sqrt{2}G}\int_{-\infty}^{T} [(\dot\sigma_R^{ijk}\sigma_R^{jk}-\sigma_R^{ijk}\dot\sigma_R^{jk})+(\dot\sigma_I^{ijk}\sigma_I^{jk}-\sigma_I^{ijk}\dot\sigma_I^{jk})]dt.
\end{equation}
Regarding the evolution of the intrinsic angular momentum, the PN approach gives a flux law for the angular momentum in the center of mass frame,
\begin{eqnarray}
\dot S^i_{PN}=&-&\epsilon^{ijk}G \Big( \frac{1}{c^5} \frac{2}{5}U^{kl}U^{(1)jl}+\frac{1}{c^5}\frac{32}{45}V^{kl}V^{(1)jl}\nonumber\\
&+&\frac{1}{c^7}\frac{1}{63}U^{klm}U^{(1)jlm} +\frac{1}{c^7}\frac{1}{28}V^{klm}V^{(1)jlm} \Big).
\end{eqnarray}
This is highly surprising since it has exactly the same r.h.s. as in eq. (\ref{impart}). It is worth making a few comments regarding the above equation. First, eq. (\ref{impart}) is derived using a specific definition of angular momentum based on linkages.  There are many formulae for angular momentum in general relativity, and all of them coincide if only quadrupole terms are taken into account. Only the linkage formulation yields the r.h.s. of eq.  (\ref{impart}).  It deserves further analysis to understand why the PN formalism yields the same r.h.s as in the linkage formula for the angular momentum loss. The second point is more subtle and deserves a closer look. It is tacitly assumed in the PN approach that the center of mass frame corresponds to a particular Bondi cut at null infinity. However, it has been shown that the intersection of the future null cone from a point in the space time with null infinity is not a Bondi cut. Thus, the l.h.s. of the above equation should not be called the time derivative of the intrinsic angular momentum. This issue can be seen more clearly in eq. (\ref{jotai}). When gravitational
radiation reaches null infinity, even if we
set $X^i=0$ the Bondi angular momentum is not equal to the intrinsic angular momentum since the cuts are different.

Thus, there is a discrepancy between the angular momentum flux formulae given by,
\begin{equation}
\dot S^k=\dot S^k_{PN} + \frac{137c^3}{168\sqrt{2}G}(\sigma_I^{jk}\sigma_R^{jki}-\sigma_I^{jki}\sigma_R^{jk})^{\cdotp}. \label{angular}
\end{equation}
Directly from (\ref{angular}) it follows that,
\begin{equation}
\Delta S^k=\Delta S^k_{PN}  + \frac{137c^3}{168\sqrt{2}G}(\sigma_I^{ij}\sigma_R^{ijk}-\sigma_I^{ijk}\sigma_R^{ij})
\end{equation}
Note that both formulations conicide up to quadrupole terms. Note also that while in the PN approach $\Delta S^k_{PN}$ does not mix different types of radiation terms, our equations contains mixed products of "electric" and "magnetic" components of the Bondi shear.

\section{Summary and Conclusions}

The purpose of this note was to compare global results coming from two completely different approaches to the motion of sources that emit gravitational radiation.

The PN approximation relies on the definition of a point particle in Newtonian mechanics and its generalization to non trivial spacetimes. The gravitational radiation is computed in a coordinate system that is well defined near the sources and it is assumed the observer is at a large but finite distance from the source.

The asymptotic formulation on the other hand, uses full knowledge of general relativity to derive exact equations of motion for global variables of an isolated system. A non trivial task is to associate these global variables to the motion of a center of mass or the time evolution of the intrinsic angular momentum. We recall that a gravitational point particle cannot be defined in general relativity. Thus, the asymptotic formulation relies on a congruence of cuts at null infinity to define a worldline in a fiducial space with a Minkowski metric.

In some sense the two formulation should help each other since they are both strong at opposite limits, one near the sources and the other one far away from them.

We ahve shown that the evolution equations for the global variables obtained in both formulations have some similarities. In fact, both formulations yield identical results if one only keeps the quadrupole mode of the radiative field. The difference arises one including octupole and higher terms in the spherical harmonic decomposition of the radiative field. It is thus, important to check if these differences are important and/or measurable. We perform a simple check using a newtonian model of two coalescing particles given in the Appendix.

Regarding the time evolution of the intrinsic angular momentum we find that they differ by a non-vanishing term, even if we time average over a period of the gravitational wave and this difference is of the same order of magnitude of the remaining terms in equation (\ref{Sdot}). Furthermore, it is not easy to see where these terms should be coming from in the PN approximation as far as the mixing between quadrupole and octupole terms is concerned.

The equations of motion for the center of mass also have, in principle, a difference between the two approaches. However, this difference might be zero or negligible for binary coalescence. If one computes this difference in newtonian mechanics for two point particles separated by a distance $r$ in the adiabatic approximation and takes a time average over a period, this difference vanishes. This follows from the formulae given in the Appendix, where the quadrupole and octupole contributions used in the PN formalism to describe black hole coalescence in circular orbits are explicitly obtained. Thus, we should not have a difference between the two formalisms when averaging over a period of the gravitational wave. We conclude that both formulations yield similar results for the center of mass motion when considering black hole coalescence.

On the other hand, gravitational waves emitted by supernovae come from a completely different physical scenario and could give different answers. If so, this could serve as a test for the formulations.

\appendix
\section{Compact Binary System} \label{Appendix}
In this appendix we derive the quadrupole and octupole moments for two spinning objects with mass $m_1$ and $m_2$ in a circular orbit in the x-y plane, at distance $r_1$, $r_2$ (respectively) from their common center of mass. The motion of the objects is considered in the Newtonian approximation.\

The mass parameters are given as $m = m_1 + m_2$, $\delta m = m_1-m_2$ and the symmetric mass ratio is given by $\eta =m_1m_2/m^2$.\

We define $\overrightarrow x=\overrightarrow r_1-\overrightarrow r_2$ and $r_s=|\overrightarrow x|$ to be the relative vector and separation between the particles.\

The motion of the two objects in the center of mass frame is equivalent to the motion of a particle of reduced mass $\mu$, the mutual action of the force that describes the mutual interaction, the force
of attraction between two masses separated by a distance $r_s = r_1 + r_2$. If this particle describes a circular motion of radius $r_s$, its acceleration is $\Omega^2r_s$. Newton second law is written.

\begin{equation}
 \mu\Omega^2r_s=\frac{Gm_1m_2}{r_s^2}
\end{equation}
and then the angular frequency of the orbit is,

\begin{eqnarray}
\Omega&=&\Big(\frac{Gm}{r_s^3}\Big)^{1/2}.
\end{eqnarray}
In terms of $\Omega$ we can write,
\begin{eqnarray}
\overrightarrow{r}_1&=&\frac{M_2}{M}r_s(\cos \Omega t, \sin \Omega t)\\
\overrightarrow{r}_2&=&-\frac{M_1}{M}r_s(\cos \Omega t, \sin \Omega t)
\end{eqnarray}

the position and relative velocity is,

\begin{eqnarray}
\overrightarrow{x}&=&\overrightarrow r_1-\overrightarrow r_2=r_s(\cos \Omega t, \sin \Omega t )\\
\dot{\overrightarrow{x}}&=&\overrightarrow v =r_s\Omega  (-\sin \Omega t, \cos \Omega t).
\end{eqnarray}

From \cite{Kidder2} o \cite{BlanchetQ}, we have the following expressions for the quadrupole and octupole moments,
\begin{eqnarray}
I^{ij}_N&=&\eta m x^{<ij>}\\
I^{ijk}_N&=&-\eta\delta m x^{<ijk>}\\
J^{ij}_N&=&-\eta\delta m \epsilon^{ab<i}x^{j>a}v^b=-\frac{\delta m}{m}L^{<i}x^{j>}\\
J^{ijk}&=&\eta(1-3\eta) m \epsilon^{ab<i}x^{jk>a}v^b=(1-3\eta)L^{<i}x^{jk>}.
\end{eqnarray}
\
In the main text of this work a comparison is made using the mass parameters of the collision of two black holes, recently detected by LIGO \cite{LIGO2016}. In this binary system the mass parameters are,
\begin{eqnarray}
M_1&=&36M_\odot\\
M_2&=&29M_\odot\\
M_F&=&62M_\odot\\
\eta&=&\frac{M_1M_2}{M^2}\approx 16\\
\delta m &=& 7M_\odot.
\end{eqnarray}
With these mass parameters, the quadrupole, octupole radiative moments are
\begin{eqnarray}
I^{ij}_N&\approx& 1040M_\odot\big[ x^ix^j-\frac{1}{3}\delta_{ij}x^2\big]\\
I^{ijk}_N&\approx&-112M_\odot \big[x^ix^jx^k-\frac{1}{5}x^2(\delta_{jk}x^i+\delta_{ik}x^j+\delta_{ij}x^k)\big]\\
J^{ij}_N&\approx&=-112M_\odot\big[\frac{1}{2}(\epsilon^{abi}x^jx^av^b+\epsilon^{abj}x^ix^av^b)-\frac{1}{3}
\delta_{ij}\epsilon^{kab}x^av^bx^k\big]  \\
J^{ijk}_N&\approx&-48880M_\odot \big[ \frac{1}{3}(L^ix^jx^k+L^jx^kx^i+L^kx^ix^j)\nonumber\\
&-&\frac{1}{15} x^2(\delta_{ij}L^k+\delta_{kj}L^i+\delta_{ik}L^j)\\
&-&\frac{2}{15} L^ax^a(\delta_{ij}x^k+\delta_{kj}x^i+\delta_{ik}x^j)\big].\nonumber
\end{eqnarray}

Explicitly the non-zero radiative moments remain,

\begin{eqnarray}
I_N^{zz}&=&-1040\frac{M_\odot r_s^2}{3}\\
I_N^{xx}&=&1040\frac{M_\odot r_s^2}{6}[1+3\cos(2\Omega t)]\\
I_N^{yy}&=&1040\frac{M_\odot r_s^2}{6}[1-3\cos(2\Omega t)]\\
I_N^{xy}&=&I_N^{yx}=1040M_\odot r^2_s[\sin\Omega t \cos\Omega t]
\end{eqnarray}

\begin{eqnarray}
I_N^{xxx}&=&-112\frac{M_\odot r_s^3}{2}\cos\Omega t[-\frac{1}{5}+\cos 2\Omega t]\\
I_N^{xxy}&=&I_N^{xyx}=I_N^{yxx}=-112\frac{M_\odot r_s^3}{10}\sin\Omega t [3+5\cos2\Omega t ]\\
I_N^{xyy}&=&I_N^{yxy}=I_N^{yyx}=-112\frac{M_\odot r_s^3}{10}\cos\Omega t [3-5\sin2\Omega t] \\
I_N^{xzz}&=&I_N^{zxz}=I_N^{zzx}=112 \frac{M_\odot r^3_s}{5}\cos\Omega t\\
I_N^{yyy}&=&112 \frac{M_\odot r^3_s}{2}\sin\Omega t[ \frac{1}{5}+\sin2\Omega t]\\
I_N^{yzz}&=&I_N^{zyz}=I_N^{zzy}=112 \frac{M_\odot r^3_s}{5}\sin\Omega t
\end{eqnarray}

\begin{eqnarray}
J_N^{xz}&=&J_N^{zx}=-112 \frac{M_\odot r_s^3}{2}\Omega\cos\Omega t\\
J_N^{yz}&=&J_N^{zy}=-112 \frac{M_\odot r_s^3}{2}\Omega\sin\Omega t
\end{eqnarray}

\begin{eqnarray}
J^{xyz}_N&=&J^{xzy}_N=J^{yxz}_N=-48880 \frac{M_\odot r_s^4}{3}\Omega\sin\Omega t \cos\Omega t\\
J^{yzx}_N&=&J^{zxy}_N=J^{zyx}_N=-48880 \frac{M_\odot r_s^4}{3}\Omega\sin\Omega t \cos\Omega t.
\end{eqnarray}

Using the above formulae and inserting the relevant terms in the center of mass equation of motion one then concludes that for this binary system both formulations yield similar results when taking an average value over a period.


\end{document}